\documentclass[reprint, aps, amsmath, amssymb, pra, notitlepage]{revtex4-1}

\usepackage{mathtools}
\usepackage{algcompatible}
\usepackage{newfloat}
\usepackage[size=small]{caption}
\usepackage{etoolbox}

\AtBeginEnvironment{algorithm}{\noindent\hrulefill\par\nobreak\vskip-5pt}
\usepackage{newfloat}

\DeclareFloatingEnvironment[
    fileext=loa,
    listname=List of Algorithms,
    name=ALGORITHM,
    placement=tbhp,
]{algorithm}
\DeclareCaptionFormat{algorithms}{\vskip-15pt\hrulefill\par#1#2#3\vskip-6pt\hrulefill}
\usepackage{graphicx}
\usepackage{dcolumn}    
\usepackage[utf8]{inputenc}
\usepackage[T1]{fontenc}
\usepackage[normalem]{ulem}
\usepackage{bm}
\usepackage{natbib}
\usepackage{amsthm}
\usepackage{epstopdf}
\usepackage{braket}
\usepackage{proof}

\usepackage{url}
\usepackage{calc}
\usepackage{tikz-qtree}
\usetikzlibrary{trees, patterns}
\usetikzlibrary{positioning}
\usepackage{tikz}
\usepackage{proof}

\usepackage{complexity}
\usepackage{fancyref}
\usepackage{lipsum}
\usetikzlibrary{shapes,snakes}
\definecolor{ltgray}{rgb}{0.95,0.95,0.95}

\newcommand{\id}{\mathbf{1}}

\usepackage{textcomp}

\newcommand{\legenddiamond}[1][fill=black]{\tikz [x=1.2ex,y=1.85ex,line width=.1ex,line join=round, yshift=-0.285ex] \node[fill,diamond, inner sep = 2pt, fill=black] at (0,0) {};}
\newcommand{\legendcircle}[1][fill=black]{\tikz [x=1.2ex,y=1.85ex,line width=.1ex,line join=round, yshift=-0.285ex] \node[fill,circle,inner sep=2.5pt,fill=black] at (0,0) {};}%
\newcommand{\legendsmallcircle}[1][fill=black]{\tikz [x=1.2ex,y=1.85ex,line width=.1ex,line join=round, yshift=-0.285ex] \node[fill,circle,inner sep=1.5pt,fill=black] at (0,0) {};}%
\newcommand{\legendsquare}[1][fill=black]{\tikz [x=1.2ex,y=1.85ex,line width=.1ex,line join=round, yshift=-0.285ex] \node[fill,rectangle,inner sep=3pt,fill=black] at (0,0) {};}%

\pgfdeclarepatternformonly{soft horizontal lines}{\pgfpointorigin}{\pgfqpoint{100pt}{1pt}}{\pgfqpoint{100pt}{3pt}}%
{
  \pgfsetstrokeopacity{0.3}
  \pgfsetlinewidth{0.1pt}
  \pgfpathmoveto{\pgfqpoint{0pt}{0.5pt}}
  \pgfpathlineto{\pgfqpoint{100pt}{0.5pt}}
  \pgfusepath{stroke}
}

\usetikzlibrary{calc,intersections, fit}

\begin{document}
 
 \title{Operator Locality in Quantum Simulation of Fermionic Models}

\begin{abstract}
Simulating fermionic lattice models with qubits requires mapping fermionic degrees of freedom to qubits. The simplest method for this task, the Jordan-Wigner transformation, yields strings of Pauli operators acting on an extensive number of qubits. This overhead can be a hindrance to implementation of qubit-based quantum simulators, especially in the analog context. Here we thus review and analyze alternative fermion-to-qubit mappings, including the two approaches by Bravyi and Kitaev and the Auxiliary Fermion transformation. The Bravyi-Kitaev transform is reformulated in terms of a classical data structure and generalized to achieve a further locality improvement for local fermionic models on a rectangular lattice. We conclude that the most compact encoding of the fermionic operators can be done using ancilla qubits with the Auxiliary Fermion scheme. Without introducing ancillas, a variant of the Bravyi-Kitaev transform provides the most compact fermion-to-qubit mapping for Hubbard-like models. 
\end{abstract}

\author{Vojtěch Havlíček}
\affiliation{Institute for Theoretical Physics and Station Q Zurich, ETH Zurich, 8093 Zurich, Switzerland}
\affiliation{Computer Science, University of Oxford, Wolfson Building, Parks Road, Oxford OX1 3QD, UK}

\author{Matthias Troyer}
 \affiliation{Institute for Theoretical Physics and Station Q Zurich, ETH Zurich, 8093 Zurich, Switzerland}
\affiliation{Quantum Architectures and Computation Group, Microsoft Research, Redmond, WA 98052, USA}
 
 \author{James~D. Whitfield}
 \affiliation{Department of Physics and Astronomy, Dartmouth College, 6127 Wilder Laboratory, Hanover, NH 03755, USA}

\maketitle
\section{Introduction} 
Among the various applications of quantum computing, quantum simulation has long stood out as a primary motivation \cite{Feynman, Lloyd}. Classical computers can often perform rapid electronic structure calculations without explicit electron-electron interaction and obtain relatively accurate results \cite{Ramakrishnan14}. However, systems where the electron-electron interaction cannot be integrated out are called strongly correlated and represent a new frontier for electronic structure in both theoretical chemistry \cite{Jiang15} and strongly correlated materials, such as high-temperature superconductors \cite{Altland2010Condensed}. It is in this regime that quantum simulation is a promising route forward \cite{Wecker, Nori, Georgescu, Dallaire16, Kreula16}.

Quantum simulation comes in two distinct flavors: digital and analog, are each subject to different mindset and constraints. In the digital context, the hardware is thought of as a universal quantum computer where an arbitrary quantum circuit can be implemented and used to approximate the system of interest \cite{Lloyd}. Since high-quality qubits are required in this context, the simulation qubit count can be thought of as an important constraint.

 An analog quantum simulator on the other hand approximates the system with another, easier to implement, control and measure. Such a simulator or emulator is usually tailored to a specific problem and it is therefore argued to be technologically more viable to build such a chip rather than a general purpose quantum computer \cite{Georgescu}. These \textit{analog} simulators are typically restricted to 2-qubit couplings and a limited set of global operations - examples being both the trapped ions \cite{Blatt} or the superconducting qubits \cite{Georgescu}.

Quantum simulation of strongly correlated fermionic systems has recently been a focus of algorithmic developments  \cite{Bela16, Kreula16, Dallarie16, Wecker}.  Besides direct simulation~\cite{Wecker}, it has been pointed out that quantum simulation of a strongly correlated region can act as an impurity solver for dynamical mean field theories \cite{Bela16, Dallaire16, Kreula16}.  In these recent investigations, the authors have chosen to use the Jordan-Wigner fermionic encoding scheme \cite{JordanWigner, Whitfield11, Somma02} for their specific simulations. However, under this transformation, local fermionic operators become spin operators acting on an extensive number of qubits, which may be problematic especially in the analog context. This can be avoided using other encoding schemes and we contribute to the ongoing line of research by investigating various fermion-to-qubit mappings.

The Hubbard model has served as a paradigmatic example for strongly correlated problems \cite{Wecker, Altland2010Condensed, Georgescu}.    We will continue this trend and use the Hubbard model as testbed for our ideas.  Its Hamiltonian, on a graph with edges $E$ and vertices $V$, is given by:
\begin{equation}
H = -t \sum_{(i,j)\in E} \sum_{\sigma=\uparrow,\downarrow} (a_{i \sigma}^{\dagger} a_{j\sigma} + a_{j \sigma}^{\dagger} a_{i\sigma} )  + U \sum_{i\in V} n_{i\uparrow} n_{i\downarrow},
\end{equation}
 where $t$ and $U$ are parameters of the model, $n_{j\sigma}=a_{j\sigma}^\dag a_{j\sigma}$, and the fermion creation operators $\{a_{i\sigma}\}$ satisfy $a_{i\sigma}^\dag a_{j\tau}+a_{j\tau} a^\dag_{i\sigma}=\delta_{ij}\delta_{\sigma\tau}$.  We will consider $t$ and $U$ to be fixed and assume, for now, that we are on a square lattice.

The paper is organized into two parts. The first part reviews and extends mappings from fermionic Hamiltonians to qubits and the second part studies operator locality of the various fermion encoding methods. 

\section{Mapping Fermionic Hamiltonians to Qubits}
The following section reviews and expands on a set of locality improving transformations for mapping fermionic Hamiltonians to qubits. 
Subsection~\ref{SSec:JW} briefly summarizes the Jordan-Wigner transformation which will be used as a baseline for locality overhead comparison.
In subsection~\ref{SSec:BK}, we review the first method originally outlined in Ref.~\cite{Kitaev} which has been referred to as Bravyi-Kitaev transformation in the literature~\cite{Seeley, Tranter}. We reformulate the transform in terms of a classical data structure (different perspective on the construction can be found in \cite{Seeley, Tranter}). This allows for its generalization outlined in subsection~\ref{SSec:FTaaCoFE}. The generalized Bravyi-Kitaev transformation corresponds to a whole class of fermion-to-spin transformations characterized by a transition from linear to logarithmic operator locality.
In part~\ref{SSec:LSFS}, we review the second method outlined in Ref.~\cite{Kitaev} and provide an example of $2$D Hubbard model mapping.
Lastly, subsection~\ref{SSec:AUX} reviews the Auxiliary Fermion method introduced in~\cite{Ball, Verstraete} with focus on operator locality analysis. We have previously worked out construction details of this transformation in Ref.~\cite{us}. 

\subsection{Jordan-Wigner Transform}
\label{SSec:JW} 
The usual way to map fermionic operators to qubits is the Jordan-Wigner (JW) transformation \cite{JordanWigner}. This encoding stores information about the occupancy of $N$ fermionic sites in $N$ qubits. The fermion raising/lowering operators on $k$-th site are mapped to qubit operators by: 
\begin{align*}
a_k &\mapsto \left( \prod_{j=0}^{k-1} Z_j \right) \, \ket{0}\bra{1}_k\,,  & a_k^\dagger &\mapsto  \left( \prod_{j=0}^{k-1} Z_j \right) \, \ket{1}\bra{0}_k \,,
\end{align*}
where $Z_j$ stands for an $N$-qubit operator corresponding to a Pauli $Z$ operator applied to the $j$-th qubit and $\id$ to the rest of the qubit register. 
The above operators obey fermionic anti-commutation relations and therefore generate fermionic algebra on qubits. The \begin{align*} \prod_{j=0}^{k-1} Z_j = \underbrace{Z \otimes \, Z \ldots \, Z}_{k} \, \otimes \underbrace{ \id \otimes \ldots \id }_{N-k} \,,\end{align*} string of Pauli $Z$ operators \textit{counts} the excitation parity. 

The action of raising/lowering operators can be therefore thought of as a composition of two operations on qubit states: (1) \textit{counting} the parity and (2) \textit{updating} the fermionic site occupancy. The number of single-qubit Pauli $Z$ operators used for parity counting scales asymptotically as $O(N)$, while the update is implemented with a single qubit operator $\ket{0} \bra{1}_k$ or its conjugate. The composite operation hence costs $O(N)$ in operator locality.

The fermionic raising/lowering operators occur only in pairs in any physical Hamiltonian. The Pauli strings could therefore cancel, as is the case in $1$D Hubbard model. In a general case however (specifically for a Hubbard model on higher dimensional lattices), the hopping operator locality scales with the size of the lattice. 
We therefore proceed by introducing an alternative scheme which improves locality of the resulting qubit Hamiltonian.

\subsection{The Bravyi-Kitaev Transform}
\label{SSec:BK}
The first of the two fermionic transformations introduced in \cite{Kitaev}, the Bravyi-Kitaev (BK) transform,  can be described by a classical data structure, the Fenwick tree \cite{Fenwick94anew}, which we will introduce below. The BK transform  has been previously reviewed in \cite{Tranter, Seeley} and formulated in terms of recursive prescription for transformation matrices. This carried an implicit constraint on the number of qubits being a power of $2$. Our approach defines the scheme for an arbitrary number of qubits. 

\subsubsection{Fenwick Trees}
In context of classical computation, a Fenwick tree can be used to map binary strings $n_0 \, n_1 \ldots n_N$, $n_i \in \lbrace 0, 1 \rbrace$ to binary strings $x_0 \, x_1 \ldots x_N$, $x_i \in \lbrace0,1\rbrace$ such that both the prefix sum $\left( \sum_{m=0}^{k-1} n_m \right)$
  and bit-flip operations have $O(\log N)$ access costs in the encoded representation. This optimization is achieved by storing partial occupancy sums ($x_i$) rather than occupancies/bits ($n_i$) in a way we now describe. The partial occupancy sums $x_i$ are dictated by the tree constructed using Algorithm~\ref{Alg:Fenwick}:\\
\begin{algorithm}
\begin{algorithmic}
\State Define \textbf{Fenwick}$(L, R)$: \\
\quad IF $L \neq R$:\\
\quad   \qquad Connect $R$ to $\lfloor \frac{R+L}{2} \rfloor$;\\
\quad \qquad  \textbf{Fenwick}({$L, \lfloor \frac{R+L}{2} \rfloor$}); \\
\quad \qquad  \textbf{Fenwick}({$\lfloor \frac{R+L}{2} \rfloor + 1, R$}); \\
\quad ELSE:\\
\quad  \qquad Terminate.
\end{algorithmic}
\caption{Fenwick Tree Generation}
\label{Alg:Fenwick}
\end{algorithm}

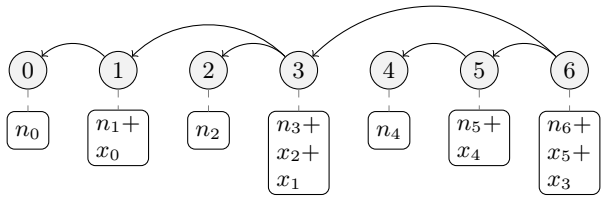
\begin{figure}[t]
\begin{tikzpicture}[every node/.style={minimum size =0.5cm}]
\node [circle, draw, inner sep = 1pt, fill = ltgray] (0) at (0,0) {0};
\node [circle, draw, inner sep = 1pt, fill = ltgray] (1) at (1.2,0) {1};
\node [circle, draw, inner sep = 1pt, fill = ltgray] (2) at (2.4,0) {2};
\node [circle, draw, inner sep = 1pt, fill = ltgray] (3) at (3.6,0) {3};
\node [circle, draw, inner sep = 1pt, fill = ltgray] (4) at (4.8,0) {4};
\node [circle, draw, inner sep = 1pt, fill = ltgray] (5) at (6,0) {5};
\node [circle, draw, inner sep = 1pt, fill = ltgray] (6) at (7.2,0) {6};

\draw [->] (6) to[out = 135, in = 45] (3);
\draw [->] (6) to[out = 135, in = 45] (5);
\draw [->] (5) to[in = 45, out = 135] (4);
\draw [->] (3) to[in=45, out=135] (2); 
\draw [->] (3) to[in=45, out=135] (1); 
\draw [->] (1) to[in=45, out=135] (0);

\node[inner sep=0pt,,outer sep=0pt,clip,rounded corners=0.1cm, draw] (sum6) at (7.2,-1.1) { $\begin{array}{l} n_6 + \\x_5 +\\ x_3 \end{array}$};
\node[inner sep=0pt,,outer sep=0pt,clip,rounded corners=0.1cm, draw]  (sum5) at (6,-.9) { $\begin{array}{l} n_5+ \\ x_4 \end{array}$};
\node[inner sep=0pt,,outer sep=0pt,clip,rounded corners=0.1cm, draw]  (sum4) at (4.8,-0.8) { $\begin{array}{l} n_4 \end{array}$};
\node[inner sep=0pt,,outer sep=0pt,clip,rounded corners=0.1cm, draw]  (sum3) at (3.6,-1.1) { $\begin{array}{l} n_3 + \\ x_2 +  \\ x_1 \end{array}$};
\node [inner sep=0pt,,outer sep=0pt,clip,rounded corners=0.1cm, draw] (sum2) at (2.4,-0.8) { $\begin{array}{l} n_2 \end{array}$};
\node[inner sep=0pt,,outer sep=0pt,clip,rounded corners=0.1cm, draw]  (sum1) at (1.2,-.9) { $\begin{array}{l} n_1+ \\ x_0 \end{array}$};
\node [inner sep=0pt,,outer sep=0pt,clip,rounded corners=0.1cm, draw] (sum0) at (0,-0.8) { $\begin{array}{l} n_0 \end{array}$};

\draw [dashed, help lines] (6) to (sum6);
\draw [dashed, help lines] (5) to (sum5);
\draw [dashed, help lines] (4) to (sum4);
\draw [dashed, help lines] (3) to (sum3);
\draw [dashed, help lines] (2) to (sum2);
\draw [dashed, help lines] (1) to (sum1);
\draw [dashed, help lines] (0) to (sum0);

\end{tikzpicture}
\caption{Fenwick tree of depth 3 for $N = 7$.  The structure can be constructed by taking the first node and making it dependent on contents of the node half way (rounded down) in the lattice and proceeding recursively for halves of the site array. The example here is illustrated for  $N = 7$. Odd $N$ has been chosen in order to show a construction of the mapping for $N$ not being a power of $2$, a restriction implicitly imposed in \cite{Seeley}. Content of the white boxes corresponds to the information stored in each node. } 
\label{Fig:Kitaev}
\end{figure}
A tree generated by $\textbf{Fenwick}(0, \, N-1)$ has depth $ d= \lceil \log_2 N \rceil $ and number of root-children equal to $n = \lfloor \log_2 N \rfloor$.  An example of a Fenwick tree for $N=7, \, d=3$ is shown in Fig.~\ref{Fig:Kitaev}. The partial sums $x_j$ of the encoded representation are given by a (mod 2) sum of $j$-th fermionic occupancy $n_j$ with the descendants of $j$ in the Fenwick tree. For example the zeroth bit encoded by a Fenwick tree in Fig.~\ref{Fig:Kitaev} stores only the occupancy of the zeroth fermionic site as it has no descendants, while the first bit stores $x_1 = n_1 +  x_0 = n_1  +  n_0$.  Likewise, the sixth bit has $\lbrace 3,5 \rbrace $ as its children and therefore stores $x_6 = n_6 + (x_3 + x_5) = n_0 + n_1 + n_2 + n_3 + n_4 + n_5 + n_6$. The remaining bits are given by:
\begin{align*}
x_0 &= n_0, & x_1 &= n_1 + x_0, & x_2 &= n_2, \\ x_3 &= n_3 + x_2 + x_1,  & x_4 &= n_4, & x_5 &= n_5 + n_4.
\end{align*}

As a specific example, $n_0\, n_1 \ldots n_6 = 0111010$ is encoded as $x_0\, x_1 \ldots x_6 = 0111010$. 

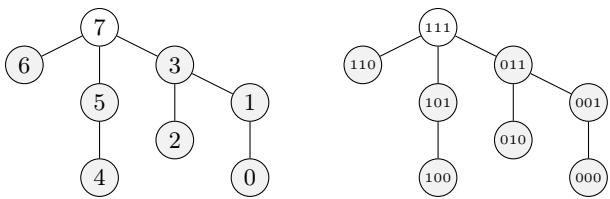
\begin{figure}[t]
\begin{tikzpicture}[every node/.style={minimum size =0.5cm}]
\node [circle, draw, inner sep = 1pt] (7) at (-2.5,-0.5) {7};
\node [circle, draw, inner sep = 1pt, fill = ltgray] (6) at (-3.5, -1){6};
\node [circle, draw, inner sep = 1pt, fill = ltgray] (5) at (-2.5, -1.5){5};
\node [circle, draw, inner sep = 1pt, fill = ltgray] (3) at (-1.5, -1){3};
\node [circle, draw, inner sep = 1pt, fill = ltgray] (4) at (-2.5, -2.5){4};
\node [circle, draw, inner sep = 1pt, fill = ltgray] (2) at (-1.5, -2){2};
\node [circle, draw, inner sep = 1pt, fill = ltgray] (1) at (-0.5, -1.5){1};
\node [circle, draw, inner sep = 1pt, fill = ltgray] (0) at (-0.5, -2.5){0};
\draw (7) -- (6);
\draw (7) -- (5) -- (4);
\draw (7) -- (3) -- (2);
\draw (3) -- (1) -- (0);

\tiny
\node [circle, draw, inner sep = 1pt] (111) at (2,-0.5) {111};
\node [circle, draw, inner sep = 1pt, fill = ltgray] (110) at (1, -1){110};
\node [circle, draw, inner sep = 1pt, fill = ltgray] (101) at (2, -1.5){101};
\node [circle, draw, inner sep = 1pt, fill = ltgray] (011) at (3, -1){011};
\node [circle, draw, inner sep = 1pt, fill = ltgray] (100) at (2, -2.5){100};
\node [circle, draw, inner sep = 1pt, fill = ltgray] (010) at (3, -2){010};
\node [circle, draw, inner sep = 1pt, fill = ltgray] (001) at (4, -1.5){001};
\node [circle, draw, inner sep = 1pt, fill = ltgray] (000) at (4, -2.5){000};
\draw (111) -- (110);
\draw (111) -- (101) -- (100);
\draw (111) -- (011) -- (010);
\draw (011) -- (001) -- (000);
\end{tikzpicture}
\caption{ Fenwick tree of depth $3$ for $N = 8$. Fenwick trees for $N=2^d$ can be also described by a partial ordering on tree node indices.  Suppose we write the indices in binary as in the tree on the right. Then a bitstring with $h > 0$ zeroes labels a child of another bitstring with $h-1$ zeroes given by flipping the last $0$ of the string to $1$. For example, $101$, $011$ and $110$ are all children of $111$. This construction manifests a possible connection to algebraic coding, as every path from the root to a leaf gives a Gray code\cite{Savage, Fenwick94anew}.  Other definitions can be found, but working out examples is the fastest way to familiarize oneself with the construction.   }
\label{Fig:TreeGray}
\end{figure}

\subsubsection{Bravyi-Kitaev Transformation}
The BK transform uses Fenwick trees to improve qubit operator locality of the fermionic parity counting string to $O(\log N)$, while increasing fermionic occupancy update cost to $O(\log N)$. The raising/lowering operators are hence mapped with $O( \log N )$ operator locality overhead, which is substantially better than $O(N)$ for JW. 

 Starting with the simplest example, consider $a^\dag_2$ applied to the second fermionic site in a \textit{qubit} register $\ket{x_0 \, x_1 \ldots x_6}$ encoding an occupancy state $\ket{n_0 \, n_1 \ldots n_6}$ of 7 fermionic sites as in Fig.~\ref{Fig:Kitaev}. This operator acts as: 
\begin{align*}
a^\dag_2 &\rightarrow Z_1 \, \ket{1} \bra{0}_2 \, X_3 \, X_6\,,
\end{align*}
on the encoded states, as one needs to \textit{count} the excitation parity of $0$ and $1$ by applying $Z_1$, change the occupancy of the second node by applying $\ket{1} \bra{0}_2$ and ensure consistency of the encoding by  \textit{updating} sites 3 and 6 (ancestors of 2) with $X_3$ and $X_6$. 

Mapping $a_j^\dag$ for a general $j$ is more complicated, as one has to condition application of $\ket{0}\bra{1}_j$ or $\ket{1}\bra{0}_j$ on content of children of $j$ in the Fenwick tree. This is the case for $j = 3$ in Fig.~\ref{Fig:Kitaev} for example. If the third fermionic site is initially unoccupied ($n_3 = 0$), the raising operator changes $n_3$ from 0 to 1. In the encoded representation, the third qubit stores $x_3 = n_0 + n_1 + n_2 + n_3 = (x_1 + x_2) + n_3$. So if $(x_1 + x_2) = 1$, a qubit lowering operator $\ket{0}\bra{1}_3$ should be applied in the encoded representation instead of $\ket{1}\bra{0}_3$. It follows that one has to condition this operation on the children's parity. The operator hence maps to:
\begin{align*}
a_3^\dag \rightarrow &-\left( \ket{1} \bra{1}_1 \ket{0}\bra{0}_2 + \ket{0} \bra{0}_1 \ket{1} \bra{1}_2 \right) \ket{0} \bra{1}_3 X_6 \\ & + \left( \ket{0} \bra{0}_1 \ket{0}\bra{0}_2 + \ket{1} \bra{1}_1 \ket{1} \bra{1}_2 \right) \ket{1} \bra{0}_3 X_6 \,.
\end{align*}

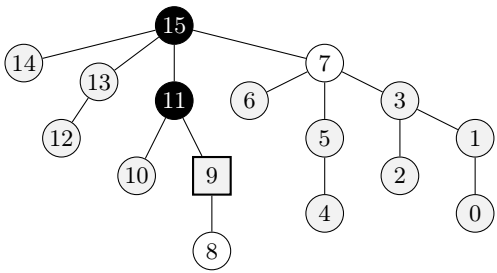
\begin{figure}[t]
\begin{tikzpicture}[every node/.style={minimum size =0.5cm}]

\node [circle, draw, inner sep = 1pt, fill = black] (15) at (0,0) {\color{white} 15};

\node [circle, draw, inner sep = 1pt, fill = ltgray] (14) at (-2,-0.5) {14};
\node [circle, draw, inner sep = 1pt, fill = ltgray] (13) at (-1,-.75) {13};
\node [circle, draw, inner sep = 1pt, fill = black] (11) at (0,-1) {\color{white} 11};
\node [circle, draw, inner sep = 1pt] (7) at (2,-0.5) {7};

\node [circle, draw, inner sep = 1pt, fill = ltgray] (12) at (-1.5,-1.5) {12};
\node [circle, draw, inner sep = 1pt, fill = ltgray] (10) at (-.5, -2){10};
\node [rectangle, draw, thick, inner sep = 1pt, fill = ltgray] (9) at (.5, -2){9};
\node [circle, draw, inner sep = 1pt, fill = ltgray] (6) at (1, -1){6};
\node [circle, draw, inner sep = 1pt, fill = ltgray] (5) at (2, -1.5){5};
\node [circle, draw, inner sep = 1pt, fill = ltgray] (3) at (3, -1){3};

\node [circle, draw, inner sep = 1pt] (8) at (.5, -3){8};
\node [circle, draw, inner sep = 1pt, fill = ltgray] (4) at (2, -2.5){4};
\node [circle, draw, inner sep = 1pt, fill = ltgray] (2) at (3, -2){2};
\node [circle, draw, inner sep = 1pt, fill = ltgray] (1) at (4, -1.5){1};

\node [circle, draw, inner sep = 1pt, fill = ltgray] (0) at (4, -2.5){0};

\draw (15) -- (14);
\draw (15) -- (13) -- (12);
\draw (15) -- (11) -- (10); 
\draw (11) -- (9) -- (8); 
\draw (15) -- (7); 
\draw (7) -- (6);
\draw (7) -- (5) -- (4);
\draw (7) -- (3) -- (2);
\draw (3) -- (1) -- (0);
\end{tikzpicture}
\caption{An example of $c_9$ Majorana operator mapped with the Bravyi-Kitaev method. The white colored nodes of the Fenwick tree correspond to $P(9) = \lbrace 7,8 \rbrace$ and are the qubits to which Pauli $Z$ operators are applied. The black nodes are in $U(9)=\lbrace 11,15 \rbrace$ to which $X$ is applied.  }
\label{Fig:c9}
\end{figure}

The above description considerably simplifies by working in the Majorana basis:
\begin{align*} c_3 =  a^\dag_3 +a_3 &\rightarrow Z_1\, Z_2 \, X_3 \, X_6 \,. \end{align*}
Consider now the set of children with indices less than $j$ of all ancestors of $j$.  We label this set as $C(j)$. For example, the set of children of all ancestors of qubit $9$ in Fig.~\ref{Fig:c9} is given by $\lbrace 7, \, 9, \,  10, \, 11, \, 13, \, 14 \rbrace$ and out of this, only $7$ is less than $9$ and hence $C(9) = \lbrace 7 \rbrace$. For consistency with refs.~\cite{Seeley, Tranter}, we denote the set of children of the $j$-th site by $F(j)$ and work with a set $P(j) = C(j) \cup F(j)$. If $U(j)$ labels the set of all ancestors of $j$, then: 
\begin{align}
c_j = a_j + a_j^\dag &\rightarrow Z_{P(j)} \, X_j \, X_{U(j)}\,,\\ 
d_j = i \left( a_j^\dag - a_j \right) &\rightarrow Z_{P(j)/F(j)} \, Y_j \, X_{U(j)} =  Z_{C(j)} \, Y_j \, X_{U(j)} \,,
\label{Eq:Majorana}
\end{align}
where $Z_{P(j)}$ implies Pauli $Z$ operators applied to qubits in a \textit{set} $P(j)$. 

Note that $P(j) \cap U(j) = \emptyset$, since all nodes in $U(j)$ have indices greater than $j$ while $P(j)$ have all indices less than $j$.
Also note that the $d_j$ operator acts trivially on the $F(j)$ qubits. Locality of the $c_j$ Majorana on qubits is hence never better than $d_j$, since no operators are applied to children of the $j$-th node (Eq.~\ref{Eq:Majorana}). The worst-case locality for the $c_j$ Majorana  operator is therefore given by $|U(j) \cup P(j)| + 1 =  |U(j)| + |P(j)| + 1$. In fact, for a Fenwick tree of $N=2^d$ sites, the locality of $c_j$ becomes \textit{exactly} $\log_2 N + 1$ as we now show: 
\begin{proof}
Let $d = 0$. Then $N = 1$ and the $c_j$ locality is $\log_21 + 1 = 1$ as there is only single node in the tree. Now suppose the locality is $\log_2 N +1$ for a tree with $N =2^d$ nodes. The $2N = 2^{d +1}$ tree is constructed by connecting roots of two $N$ trees - compare for example the descendants of $7$ to the tree of the remaining nodes in Fig.~\ref{Fig:WorstCase}.
Any node in the right subtree has to update the new root, which worsens the locality by $1$ - hence the terms are now $\log_2 2N + 1$ local. Operators on the rest of the tree will have to lookup an additional node to obtain the parity, which also implies  $\log_2 2N + 1$ locality. So every $c_j$ on a tree with $N = 2^d$ nodes is $\log_2 N + 1$ local.
\end{proof}

We now provide a unifying framework for JW and BK encodings and introduce an optimized variant of the method suitable for rectangular lattice geometries. 

\subsection{Fenwick Trees as a Class of Fermionic Encodings}
\label{SSec:FTaaCoFE}
\begin{figure}[th]
\begin{tikzpicture}[every node/.style={minimum size =0.5cm}]
\node [circle, draw, inner sep = 1pt, fill = ltgray] (0) at (0,0) {0};
\node [circle, draw, inner sep = 1pt, fill = ltgray] (1) at (.75,0) {1};
\node [circle, draw, inner sep = 1pt, fill = ltgray] (2) at (1.5,0) {2};
\node [circle, draw, inner sep = 1pt, fill = ltgray] (3) at (2.25,0) {3};
\node [circle, draw, inner sep = 1pt, fill = ltgray] (4) at (3,0) {4};
\node [circle, draw, inner sep = 1pt, fill = ltgray] (5) at (3.75,0) {5};
\node [circle, draw, inner sep = 1pt, fill = ltgray] (6) at (4.5,0) {6};
\node [circle, draw, inner sep = 1pt, fill = ltgray] (7) at (5.25,0) {7};

\node [circle, draw, inner sep = 1pt, fill = ltgray] (0) at (0,1.25) {0};
\node [circle, draw, inner sep = 1pt, fill = ltgray] (1) at (.75,1.25) {1};
\node [circle, draw, inner sep = 1pt, fill = ltgray] (2) at (1.5,1.25) {2};
\node [circle, draw, inner sep = 1pt, fill = ltgray] (3) at (2.25,1.25) {3};
\node [circle, draw, inner sep = 1pt, fill = ltgray] (4) at (3,1.25) {4};
\node [circle, draw, inner sep = 1pt, fill = ltgray] (5) at (3.75,1.25) {5};
\node [circle, draw, inner sep = 1pt, fill = ltgray] (6) at (4.5,1.25) {6};
\node [circle, draw, inner sep = 1pt, fill = ltgray] (7) at (5.25,1.25) {7};
\draw[<-] (0) to[in=135, out=45]  (1) ;
\draw[<-] (2) to[in=135, out=45]  (3) ;
\draw[<-] (4) to[in=135, out=45]  (5) ;
\draw[<-] (6) to[in=135, out=45]  (7) ;

\node [circle, draw, inner sep = 1pt, fill = ltgray] (0) at (0,2.5) {0};3
\node [circle, draw, inner sep = 1pt, fill = ltgray] (1) at (.75,2.5) {1};
\node [circle, draw, inner sep = 1pt, fill = ltgray] (2) at (1.5,2.5) {2};
\node [circle, draw, inner sep = 1pt, fill = ltgray] (3) at (2.25,2.5) {3};
\node [circle, draw, inner sep = 1pt, fill = ltgray] (4) at (3,2.5) {4};
\node [circle, draw, inner sep = 1pt, fill = ltgray] (5) at (3.75,2.5) {5};
\node [circle, draw, inner sep = 1pt, fill = ltgray] (6) at (4.5,2.5) {6};
\node [circle, draw, inner sep = 1pt, fill = ltgray] (7) at (5.25,2.5) {7};
\draw[<-] (0) to[in=135, out=45]  (1) ;
\draw[<-] (2) to[in=135, out=45]  (3) ;
\draw[<-] (4) to[in=135, out=45]  (5) ;
\draw[<-] (6) to[in=135, out=45]  (7) ;
\draw[<-] (5) to[in=135, out=45]  (7) ;
\draw[<-] (1) to[in=135, out=45]  (3) ;

\end{tikzpicture}
\caption{Recursion steps of the Fenwick tree algorithm shown for $N=8$.  The bottom case of  corresponds to the JW transformation.}
\label{Fig:FenIter}
\end{figure}
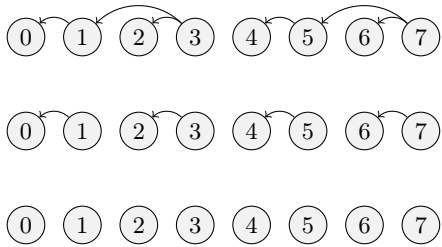

The recursive Algorithm~\ref{Alg:Fenwick} from the previous section gives rise to a class of fermionic encodings with JW and BK schemes as limiting cases. Instead of using a single Fenwick tree to encode all fermionic modes, we partition the fermionic sites into Fenwick trees of varying depth and include the set of all roots of segmented trees less than $j$ to $P(j)$. The definition of $P(j)$ in this context becomes: 
 \begin{align*} P(j) &= F(j) \cup C(j) \cup \lbrace \text{set of all roots $i$, $i < j$} \rbrace \,. \end{align*}   
 In particular, we could choose to put each node in its own Fenwick tree of depth 0, which would correspond directly to the JW transformation (Fig.~\ref{Fig:FenIter}). 
 
Every iteration of the algorithm defines new representation of fermionic algebra with qubit operators. With increasing recursion depth, the occupancy update cost worsens, while the parity counting costs improve. Asymptotically, the representation transitions from operator locality overheads of $O(1) \rightarrow O(\log N)$ for occupancy update and $O(N) \rightarrow O(\log N)$ for parity computation.

\subsection{Segmented Bravyi-Kitaev (SBK) transform}  
\label{Subsec:Opti}
It is possible to use the segmented transform for operator locality improvement of specific fermionic Hamiltonians on rectangular lattices. 
As already discussed, the BK transform optimizes update and lookup costs simultaneously. Composition of these operations describes the action of fermionic raising/lowering operators, but does not strictly correspond to operators occurring in the Hubbard model - or in fact any other physical fermionic Hamiltonian.  For closed systems, physical fermionic Hamiltonians only contain \textit{pairs} of raising and lowering operators \cite{Kitaev}, which places further constraints on the set of operators we need to map to qubits. If we additionally restrict our attention to operators which are local (as is the Hubbard model), there is a lot of redundancy one can exploit for further qubit operator locality optimization.

We focus on the case of a $w \times h, \; w \leq h$ rectangular lattice and build a Fenwick tree for every row - this is a specific example of segmented Fenwick tree as defined in the previous section. At first sight, this appears to worsen locality of the qubit operators as the locality of single raising/lowering operators now scales asymptotically as $O ( h \log w )$. If we however restrict the set of possible operations to on-site and nearest-neighbor terms, the single-qubit operators on segmented tree roots cancel and the locality becomes $O( \log w) $ - a substantial improvement, as the operator locality is now independent of the lattice height.

\begin{figure}[t]
\includegraphics[scale=0.7]{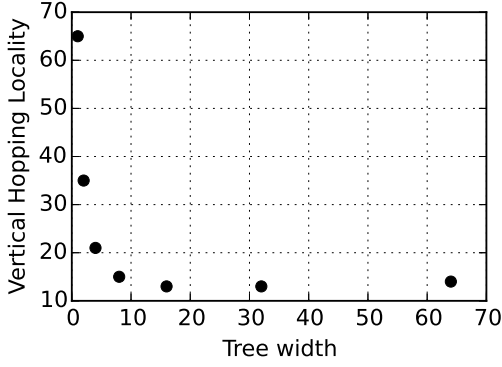}
\caption{Locality of vertical hopping terms as a function of the segment tree size for $W = 64$. The optimal tree size is $W/2$. The hopping term localities of segments of $W=64$ and $W=32$ sizes are given by $14$ and $13$ in this lattice.} 
\label{fig:vert_locality}
\end{figure}

It turns out to be slightly more optimal to store two disconnected trees per lattice row, as shown in Fig.~\ref{fig:vert_locality}. This is because the full parity of the row is not necessary for the vertical nor horizontal hopping terms.

\subsection{Loop-Stabilized Fermion Simulation (LSFS)  }
\label{SSec:LSFS}
\begin{figure}[t]
\centering
\begin{tikzpicture}
\begin{scope}

\draw[style=help lines,dashed] (0,0) grid[step=1cm] (3,3); 
\tiny 
\foreach \x in {0,1,...,3}{                           
    \foreach \y in {0,1,...,3}{                       
    \pgfmathparse{int(4*(3-\y)+ \x)}\let \j\pgfmathresult
    \node[fill=ltgray,circle,inner sep=0,draw, minimum size = 0.3cm ] at (\x, \y) { $\j$}; 
    \ifthenelse{\x < 3.5}{\node[fill,circle,inner sep=1pt,fill=black] at (\x + 0.5, \y) {}}{}; 
    \ifthenelse{\y < 3.5}{\node[fill,circle,inner sep=1pt,fill=black] at (\x,0.5+\y) {}}{};
    }
}
\end{scope}
\begin{scope}
\fill[fill=ltgray] (5.5,1) rectangle (6.5,2);
\tiny
\draw[style=help lines,dashed] (0,0) grid[step=1cm, shift={(4.5,0)}] (3,3); 
\foreach \x in {0,1,...,3}{                          
    \foreach \y in {0,1,...,3}{                       
     \pgfmathparse{int(4*(3-\y)+ \x)}\let \j\pgfmathresult
      \ifthenelse{\NOT \j = 6 \AND \NOT \j = 10 \AND \NOT \j = 5 \AND \NOT \j = 9}
    		{\node[fill=ltgray,circle,inner sep=0,draw, minimum size = 0.3cm ](\j) at (4.5+\x,\y) { $\j$}}
    		{\node[fill=gray,circle,inner sep=0,draw, minimum size = 0.3cm ](\j) at (4.5+\x,\y) { $\j$}}; 
      \ifthenelse{\x < 3.5}{\node[fill,circle,inner sep=1pt,fill=black] at (\x + 5, \y) {}}{}; 
    \ifthenelse{\y < 3.5}{\node[fill,circle,inner sep=1pt,fill=black] at (4.5+\x,0.5+\y) {}}{};
    }
}

\node[circle,inner sep=2pt,fill=black ] at (6,1) {};
\node[circle,inner sep=2pt,fill=black] at (6, 2) {};
\node[diamond, inner sep = 1.6pt, fill=black] at (6.5, 1.5) {};
\node[diamond, inner sep = 1.6pt, fill=black] at (5.5, 1.5) {};

\node[rectangle, fill=white, draw, inner sep = 2pt, minimum size = 0.3cm ] at (7, 2){Z};
\node[rectangle, fill=white, draw, inner sep = 2pt, minimum size = 0.3cm ] at (5, 1){Z};
\node[rectangle, fill=white, draw, inner sep = 2pt, minimum size = 0.3cm ] at (6, 2){X}; 
\node[rectangle, fill=white, draw, inner sep = 2pt, minimum size = 0.3cm ] at (6, 1){X}; 
\node[rectangle, fill=white, draw, inner sep = 2pt, minimum size = 0.3cm ] at (6.5, 1.5){Y}; 
\node[rectangle, fill=white, draw, inner sep = 2pt, minimum size = 0.3cm ] at (5.5, 1.5){Y}; 
\end{scope}
\end{tikzpicture}
\caption{(left) Fermionic site ordering used with the BK scheme. Qubits ($\legendsmallcircle$) are on edges of the lattice. 
(right) A stabilizer on $(5,9,6,10)$ plaquette. There are $9$ such stabilizers for $4\times 4$ lattice corresponding to the 9 plaquettes. If a state is a simultaneous eigenstate of all stabilizers, it encodes a physical fermionic state. If a qubit operator is to be applied to the edge outside the lattice, ignore it.}
\label{fig:ordering}
\end{figure}
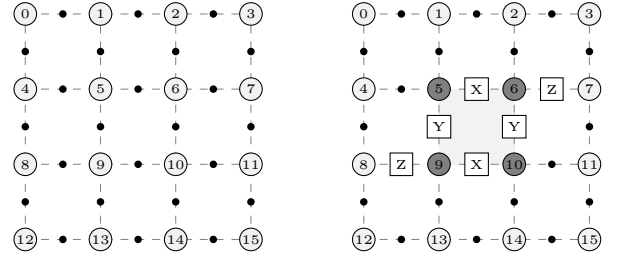

We now shift our attention to the second method introduced in Ref.~\cite{Kitaev} which finds an alternative representation of the local fermionic Hamiltonians on qubits in a line graph (i.e. on graph edges - see Fig.~\ref{fig:ordering}). The method, referred to as a ``Superfast Simulation of Fermions'' in~\cite{Kitaev} improves qubit operator locality to a constant for any bounded-degree graph. The newly proposed name ``Loop-Stabilized Fermion Simulation'' is motivated by the fermionic algebra being represented in a subspace defined by a stabilizer condition on the set of all possible loops in the line graph. 

\subsubsection{Loop-Stabilized Fermionic Simulation}
As discussed in the construction of the SBK method, physical \textit{fermionic} Hamiltonians are sums or products of fermionic raising/lowering operator pairs. Equivalently, all fermionic Hamiltonians can be obtained by combining the following operators \cite{Kitaev}:
\begin{align*}
B_k & 
= 1 - 2a_k^\dagger a_k &&\text{for a vertex } k \,, \\
 A_{(jk)} &= -i(a_{j} + a_j^\dagger)(a_k + a_k^\dagger) &&\text{for an edge } (j,k) \,. 
\end{align*} 
Here, the subscript on $B$ corresponds to a vertex and the subscript on $A$ labels a graph edge. Specifically, the fermionic site to site hopping is expressed by:
\begin{align*}
a_k^\dagger a_j + a_j^\dagger a_k  &= -i\left( A_{(jk)} B_k + B_j A_{(jk)} \right)/2\,.
\end{align*}
 More formally, the $A_{(jk)},\,B_j$ operators generate the algebra of physical fermionic Hamiltonians. The algebra defining rules are~\cite{Kitaev}: 
\begin{align*}
 A_{(jk)} B_l &= (-1)^{\delta_{jl} + \delta_{kl}} B_l A_{(jk)}, &  [B_k, B_l] &= 0, \\  A_{(jk)} A_{(ls)} &= (-1)^{\delta_{jl} + \delta_{js} + \delta_{kl} + \delta_{ks}} A_{(ls)} A_{(jk)}\,,
\end{align*}
 In other words, $A_{(jk)}$ anticommutes with any other generator as long as they share precisely one vertex. 
Furthermore note that $B_k^\dag = B_k$, $A_{(ij)}^\dag = A_{(ij)}$, $B_k^2 = A_{(ij)}^2 = \id$ and $A_{(ij)} = -A_{(ji)}$. Additionally:  
\begin{align}
(i)^p A_{(j_0 j_1)} A_{(j_1 j_2)} \ldots A_{(j_p j_0)} &= \id\,,
\label{Eq:loop}
\end{align}
for any closed path $j_0 \, j_1  \, j_2 \ldots \, j_p$.

Bravyi and Kitaev found in \cite{Kitaev} qubit operators $\tilde{B}_l, \, \tilde{A}_{(jk)}$ obeying the above algebra, subject to further constraints on excitation parity sector that we now impose. For an \textit{even} number  $N_f$ of fermions, one has that:
 \begin{align*} \prod_{k \in V} B_k &= \prod_{k \in V} (\id_k - 2a_k^\dag a_k)  = (-1)^{N_f}\id = \id\,,\end{align*} as an additional rule to the above set.
This motivates the following choice for the qubit operator $\tilde B_k$:
$$\tilde B_{k}=\prod_{j \in n(k)} Z_{(jk)}\,,$$ where $n(k)$ is the set of nearest neighbors of $k$.
The $\prod_{k \in V} \tilde{B}_k~=~\id$ condition is then trivially satisfied for the operator since each edge shares precisely two vertices (this is colloquially known as the ``handshaking lemma''). 
  
We now derive the form of $\tilde{A}_{(jk)}$. Firstly assume that $\tilde{A}_{(jk)}$ is a tensor product of Pauli operators and/or identity on edges adjacent to vertices $j,\, k$.  In order to satisfy $A_{(jk)} B_l = (-1)^{\delta_{jl} + \delta_{kl}} B_l A_{(jk)}$, we first address the case for which $k \neq l, \, j \neq l$, so that the fermionic operators commute.  If $\tilde{A}$ and $\tilde{B}$ are to obey the same relation, all single qubit operators of $A_{(jk)}$ on edges adjacent to $j, \, k$, except for the edge $(j,k)$, must be Hermitian and identity-squaring operators in a subspace spanned by $\lbrace \id, Z \rbrace$. There are only two options - either $Z$ \textit{or} $\id$. Focusing on the case $l \in \lbrace j,k\rbrace$, the $A_{(jk)} B_l$ operators anti-commute, implying that the operator on the $(j,k)$ edge qubit is a Hermitian and identity-squaring operator in the subspace spanned by $\lbrace X, Y \rbrace $ - again $X$ \textit{or} $Y$ are the only possibilities.

\begin{figure}[tb]
    \begin{tikzpicture}
\begin{scope}

\draw[style=help lines,dashed] (0,0) grid[step=1cm] (3,3); 
\tiny 
\foreach \x in {0,1,...,3}{                           
    \foreach \y in {0,1,...,3}{                       
    \pgfmathparse{int(4*(3-\y)+ \x)}\let \j\pgfmathresult
    \ifthenelse{\NOT \j = 9 \AND \NOT \j = 10}
    		{\node[fill=ltgray,circle,inner sep=0,draw, minimum size = 0.3cm ](\j) at (\x,\y) { $\j$}}
    		{\node[fill=gray,circle,inner sep=0,draw, minimum size = 0.3cm ](\j) at (\x,\y) { $\j$}}; 
    \ifthenelse{\x < 3.5}{\node[fill,circle,inner sep=1pt,fill=black] at (\x + 0.5,\y) {}}{}; 
    \ifthenelse{\y < 3.5}{\node[fill,circle,inner sep=1pt,fill=black] at (\x, 0.5+\y) {}}{};
    }
}

\node[rectangle, fill=white, draw, inner sep = 2pt, minimum size = 0.3cm ] at (1, 1.5){Z};
\node[rectangle, fill=white, draw, inner sep = 2pt, minimum size = 0.3cm ] at (.5,1){Z}; 
\node[rectangle, fill=white, draw, inner sep = 2pt, minimum size = 0.3cm ] at (1.5, 1){X}; 
\node[rectangle, fill=white, draw, inner sep = 2pt, minimum size = 0.3cm ] at (2, 1.5){Z}; 
\end{scope}
\begin{scope}

\draw[style=help lines,dashed] (5,0) grid[step=1cm, shift={(-.5,0)}] (8,3); 
\tiny 
\foreach \x in {0,1,...,3}{                          
    \foreach \y in {0,1,...,3}{                       
    \pgfmathparse{int(4*(3-\y)+ \x)}\let \j\pgfmathresult
    \ifthenelse{\NOT \j = 6 \AND \NOT \j = 10}
    		{\node[fill=ltgray,circle,inner sep=0,draw, minimum size = 0.3cm ](\j) at (\x + 4.5,\y) { $\j$}}
    		{\node[fill=gray,circle,inner sep=0,draw, minimum size = 0.3cm ](\j) at (\x + 4.5,\y) { $\j$}}; 
    
    \ifthenelse{\x < 3.5}{\node[fill,circle,inner sep=1pt,fill=black] at (\x + 5,\y) {}}{}; 
    \ifthenelse{\y < 3.5}{\node[fill,circle,inner sep=1pt,fill=black] at (4.5+ \x,0.5+\y) {}}{};
    }
}
\node[rectangle, fill=white, draw, inner sep = 2pt] at (6, 2){Z};
\node[rectangle, fill=white, draw, inner sep = 2pt] at (7,2){Z}; 
\node[rectangle, fill=white, draw, inner sep = 2pt] at (6.5, 2.5){Z}; 
\node[rectangle, fill=white, draw, inner sep = 2pt] at (6.5, 1.5){X}; 
\end{scope}
\end{tikzpicture}
\caption{Example of LSFS generators $A_{(9, 10)}$ and $A_{(6, 10)}$. }
\label{Fig:KitaevOps}
\end{figure}
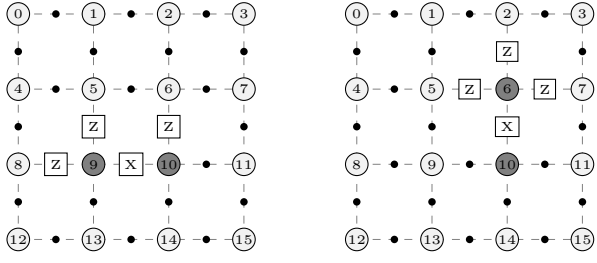

It remains to satisfy the condition imposed by the generalized commutator of two $A_{(jk)}, A_{(lm)}$ operators. These anticommute iff they share a vertex. An example of qubit operators $\tilde{A}_{(jk)}$ satisfying this and the previous constraints on a square lattice is given in Fig.~\ref{Fig:KitaevOps}. 
An example of operators satisfying these constraints on a general lattice is given by:  
\begin{align*}
\tilde{A}_{(jk)} &\propto X_{(jk)} \prod_{l<k}^{n(j)} Z_{(lj)} \prod_{s<j}^{n(k)} Z_{(sk)}\,,
\end{align*} 
and is a specific case of the prescription found in Ref.~\cite{Kitaev}. Lastly, we impose anticommutation of $A_{(jk)}$ using  antisymmetric tensor $\epsilon_{jk}$ which is +1 when $j>k$ and $-1$ when $j<k$. Thus,
\begin{align*}
\tilde{A}_{(jk)} = \epsilon_{jk} X_{(jk)} \prod_{l<k}^{n(j)} Z_{(lj)} \prod_{s<j}^{n(k)} Z_{(sk)}\,.
\end{align*}
Even though this implies the generator itself is multiply defined for each edge, we will see that all interesting physical operators will be independent of $\epsilon_{jk}$. 

The loop condition of Eq.~\ref{Eq:loop} will be imposed by a set of stabilizer operators which can be concisely presented with a specific lattice geometry in mind. We therefore postpone its discussion until after the following example. 
\subsubsection{Example: 2D Hubbard Model}
We illustrate this method by mapping the Hubbard Hamiltonian on rectangular lattice to qubits. Let $Z^\uparrow_k$ denote a Pauli $Z$ operator applied to the qubit on the vertical edge adjacent to the vertex $k$ - if there is no such edge, substitute the term with an identity operator (for example, $Z_1^\uparrow = \id$ in the diagram in Fig.~\ref{fig:ordering}). Operators on other adjacent edges are defined analogously by using $\lbrace \rightarrow, \leftarrow, \uparrow, \downarrow \rbrace$ superscripts. 

 The $\tilde{B}_k$ operator is represented with a ``cross'' of Pauli $Z$ operators:  
\begin{align*}
\tilde{B}_k &= \prod_{j \in n(k)} Z_{jk} = Z_k^\leftarrow Z_k^\uparrow Z_k^\rightarrow Z_k^\downarrow\,, & \text{for a vertex}\; k\,.
\end{align*}

The form of $\tilde{A}_{(jk)}$ differs for horizontal and vertical edges, which we denote by $E_H$ and $E_V$ respectively. It also depends on a specific lattice indexing - we choose the simplest one shown in Fig.~\ref{fig:ordering} and leave it an open question whether a better ordering exists. The operator $\tilde{A}_{(jk)}$ is then given by: 
\begin{align*}
\tilde{A}_{(jk)} &= \begin{cases} \epsilon_{jk} X_{jk} Z^\leftarrow_j Z_j^\uparrow Z_j^\rightarrow &\text{for}  \, (j,k) \, \in \, E_V \,, \\
  \epsilon_{jk} X_{jk} Z_j^\uparrow Z_k^\uparrow Z_j^\leftarrow & \text{for}  \, (j,k) \, \in \, E_H \,. \end{cases}
\end{align*}

It remains to satisfy the loop condition of Eq.~\ref{Eq:loop}, since in order to represent the fermionic algebra, the qubit operators must obey the same relation. Because the fermionic operators obey:
\begin{align*}
 A_{(\alpha \beta)} A_{(\beta \gamma)} A_{(\gamma \delta)} A_{(\delta \alpha)} &= (-i)^4 c_\alpha c_\beta c_\beta c_\gamma c_\gamma c_\delta c_\delta c_\alpha = \id\,, \\ & c_\alpha = \left( a_\alpha + a^\dag_\alpha \right)\,,
 \end{align*}
for any plaquette $(\alpha \beta \gamma \delta)$ and an arbitrary closed loop of Eq.~\ref{Eq:loop} can be obtained by taking product of such plaquette operators, encoded physical fermionic states correspond to qubit states which are $+1$ eigenstates of $\tilde{A}_{(\alpha \beta)} \tilde{A}_{(\beta \gamma)} \tilde{A}_{(\gamma \delta)} \tilde{A}_{(\delta \alpha)}$.
Formally, we restrict the qubit states by a set of \textit{stabilizer} operators: 
\begin{align*}
C_{(\alpha \beta \gamma \delta )} &=  \, \tilde{A}_{(\alpha \beta)} \tilde{A}_{(\beta \gamma) } \tilde{A}_{(\gamma \delta)} \tilde{A}_{(\delta \alpha)} \,,
\end{align*} 
where $(\alpha \beta \gamma \delta)$ labels vertices of the plaquette. An example of such stabilizer is worked out in Figure~\ref{fig:ordering} for a trivial lattice ordering. Since $C_{(\alpha \beta \gamma \delta)}$ is a Kr\"{o}necker product of Pauli matrices, its eigenvalues are $\pm 1$, implying that violation of a single stabilizer condition costs $2$ units. Note that if the simulation starts in the subspace encoding physical fermionic states, it stays in it, as the generating operators  $B_k, \, A_{(jk)}$ commute with the stabilizers \cite{Kitaev}. 

Inspired by the toric code construction~\cite{Toric}, we note that one can restrict the system to the physical codespace by including a penalty term $H_{penalty} = - \frac{\Delta}{2} \sum_{k}C_k$, where $\Delta \gg t,U,\epsilon$ corresponds to the ``energy gap'' of the system and $k$ runs over all plaquettes.

The above representation is now applied to the fermionic nearest-neighbor hopping operator. 
The nearest-neighbor couplings for horizontal edges maps to $5$-qubit-local operators: 
\begin{align*}
a_{k+1}^\dagger a_k + a_k^\dagger a_{k+1} &\rightarrow  \frac{1}{2}  Y_k^\rightarrow ( Z_k^\downarrow Z_{k+1}^\uparrow - Z_k^\uparrow Z_k^\leftarrow Z_{k+1}^\rightarrow Z_{k+1}^\downarrow) \,.
\end{align*}
Analogously, the vertical nearest-neighbor couplings are encoded by 7-local operators: 
\begin{align*}
a_{j}^\dagger a_k + a_k^\dagger a_{j} &\rightarrow \frac{1}{2} Y_j^\uparrow \left( Z_k^\leftarrow Z_k^\rightarrow Z_k^\uparrow  Z_j^\leftarrow Z_j^\rightarrow Z_j^\downarrow -  \id \right)\,.
\end{align*}

It remains to account for the Hubbard repulsion term. The simplest way to implement it is by simulating the above on two lattices labeled with $\uparrow, \downarrow$, coupled by the density-density interaction term. We obtain that:
\begin{align*}
H_\uparrow &= -t \sum_{(i,j) \in E } (a_{i\uparrow}^\dagger a_{j\uparrow} + a_{j\uparrow}^\dagger a_{i\uparrow} )  + \epsilon \sum_{i \in V}   a_{i\uparrow}^\dagger a_{i\uparrow} \,, \\
H_\downarrow &= -t \sum_{(i,j) \in E } (a_{i\downarrow}^\dagger a_{j\downarrow} + a_{j\downarrow}^\dagger a_{i\downarrow} )  + \epsilon \sum_{i \in V}   a_{i\downarrow}^\dagger a_{i\downarrow}  \,, 
\end{align*}
which leads to the following expression for the full Hamiltonian: 
\begin{align*}
H &= H_\uparrow + H_\downarrow + U \sum_{i \in V} n_{i \uparrow} n_{i\downarrow}\,.
\end{align*}
The $H_\uparrow$ and $H_\downarrow$ Hamiltonian terms are decoupled and have been already mapped to spins. The density-density interaction term maps to: 
\begin{align*}
 n_{k \uparrow} n_{k\downarrow} &\rightarrow \frac{1}{4} ( \id  - Z_k^\leftarrow Z_k^\uparrow Z_k^\rightarrow Z_k^\downarrow)  ( \id  - Z_{k'}^\leftarrow Z_{k'}^\uparrow Z_{k'}^\rightarrow Z_{k'}^\downarrow) \,, 
\end{align*}
where the primed indices correspond to fermions in spin $\downarrow$ lattice and the unprimed ones to the sites in the spin $\uparrow$ lattice. It hence follows that in order to implement this inter-lattice coupling on spins, the density-density Hamiltonian term acts on 8 qubits simultaneously in the worst case. This presents an upper bound on the spin Hamiltonian locality in this setting. 

\subsection{The Auxiliary Fermion Scheme}
\label{SSec:AUX}
The auxiliary fermion (AF) scheme \cite{us, Ball, Verstraete} was introduced in \cite{Ball} and also (independently) by \cite{Verstraete}.  
The scheme uses auxiliary fermionic modes to allow 
fermionic models on general lattice geometries 
to be simulated 
locally. We recall the relevant details from \cite{us} to allow comparison against the other fermionic encoding schemes.

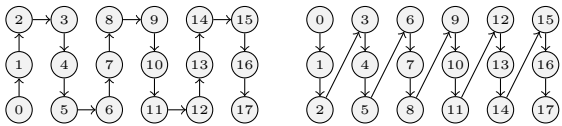
\begin{figure}[t]
\begin{tikzpicture}
\begin{scope}
\tiny
\foreach \x in {0,1,...,2}{                           
    \foreach \y in {0,1,...,5}{                      
     \pgfmathparse{int(3*(\y)+ \x)}\let \j\pgfmathresult
     \pgfmathparse{int(3*(\y)- \x+2)}\let \k\pgfmathresult
     \pgfmathparse{int(Mod(int(\y+1),2))}\let \b \pgfmathresult
     \ifthenelse{\b = 1}
        		{\node[fill=ltgray, draw, circle, minimum size=0.35cm, inner sep=0pt](\j) at (.6*\y,.6*\x) { $\j$}}
    		{\node[fill=ltgray, draw, circle, minimum size=0.35cm, inner sep=0pt](\k) at (.6*\y,.6*\x) { $\k$}}; 
     }
}

\draw [->] (0) edge (1) (1) edge (2) (2)  edge (3) (3) edge (4) (4) edge (5) (5) edge (6) (6)  edge (7) (7) edge (8);
\draw [->] (8) edge (9) (9) edge (10) (10)  edge (11) (11) edge (12) (12) edge (13) (13) edge (14) (14) edge (15) (15) edge (16) (16) edge (17);
\end{scope}

\begin{scope}
\tiny
\foreach \x in {0,1,...,2}{                           
    \foreach \y in {0,1,...,5}{                       
    \pgfmathparse{int(8 - (3*(2-\y)+ \x))}\let \j\pgfmathresult
    \node[fill=ltgray, draw, circle, minimum size=0.35cm , inner sep = 0pt](\j) at (4 + .6*\y, 0.6*\x) { $\j$}; 
    }
}
\draw [->] (0) edge (1) (1) edge (2) (2)  edge (3) (3) edge (4) (4) edge (5) (5) edge (6) (6)  edge (7) (7) edge (8) (8) edge (9);
\draw [->] (9) edge (10) (10) edge (11) (11)  edge (12) (12) edge (13) (13) edge (14) (14) edge (15) (15) edge (16) (16) edge (17);
\end{scope}
\end{tikzpicture}
\caption{(left) Snake pattern used for $G_1$ in a $w=l=3$ in the auxiliary fermion method. (right) Optimal JW ordering for geometrically local fermionic models. }
\label{Fig:snake}
\end{figure}

A given set of site-to-site hopping terms can be characterized by a graph $G=(E,\,V)$, where vertices correspond to sites and the edges to pairs of sites participating in the hopping.
In one dimension, this graph is a linear path, $G_1$, where the degree of the endpoints is one and all other sites have degree two.  In higher dimensional settings such as the $2$D Hubbard model studied here, the degree of the sites may be greater than two.  The non-local degree $d_{nl}$ of a site is the number of edges that are not included in the linear path $G_1$.  
Therefore, it is important to choose a path which overlaps maximally with the desired interaction graph $G$. In the case of $2$D Hubbard, this is accomplished by a snake-like pattern (Fig.~\ref{Fig:snake}).
Next, we must account for spin in the Hubbard model.   One requires $2wh$ spinless fermionic sites if the model is defined on $w \times h$ rectangular lattice.  If we take the first half of these sites to be spin down and the second half to be spin up, there is no need to track phase factors between the spin up and spin down subsets.  This is because the Hubbard model preserves spin; thus, there are no hopping terms between the first half of the sites and the second half. The density-density terms are each a product of two one-point coupling terms which does not require the tracking of antisymmetric phase factors.

The nonlocal degree of each node (when $G_1$ is a subgraph of $G$) is $d_{nl}=d(G)-d(G_1)$ and, as we've previously shown, each auxiliary fermionic site can facilitate up to two non-local couplings \cite{us}.
Hence the number of auxiliary sites required per fermionic site is given by  $\lceil d_{nl}/2 \rceil$.
Since each non-local degree in the $2$D Hubbard model is less than or equal to two, only a single auxiliary mode 
is needed for each site with non-local degree greater than zero. 
The $(1,w)$ and $(h,1)$ corners (for each spin) require no auxiliary sites since the edge set of $G_1$
is sufficient for coupling them to their nearest neighbors.  

As an example, we illustrate these ideas using a square lattice of 
width $w=3$ and height $h = 3$. In the general case, there are $(w-2)(h-2)$ sites with degree four 
in the interior of the graph $G$.  Each of these sites has non-local degree two and hence
each needs just a single auxiliary fermionic site.  Then there are $2(h-2)+2(w-2)$ sites along the
boundary with degree three.  Each of these sites have non-local degree of 1 and will also require
one auxiliary fermionic site each.  Finally, the four sites occupying the corners of the lattice
to be simulated have degree two.  Presuming that the snake-like pattern is used, two of the corners
will have non-local degree zero while the other two each have non-local degree 1.  Summing all the
cost together we find that the total number of qubits is $4wh-4$.

The locality of simulation operators was detailed in our previous publication \cite{us}. For a hopping term
from site $i$ to a non-consecutive site $j$, the qubit operator acts on site $i$ and on site $j$ as well as
auxiliary modes associated with site $j$ and with site $i$.  Hence, the qubit operator is either
$4$-local or $3$-local in the Hubbard model.  The density-density operators are $2$-local \cite{us}.

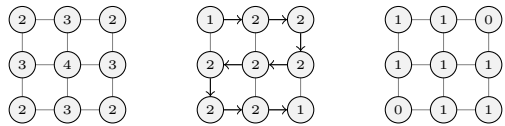
\begin{figure}[t]
\begin{tikzpicture}
\begin{scope}
\draw[help lines] (0,0) grid[step=0.6cm] (1.2,1.2); 
\def\myArray{{2,3,2,3,4,3,2,3,2}}
\tiny
\foreach \x in {0,1,2}{                           
    \foreach \y in {0,1,2}{                       
    \pgfmathtruncatemacro{\j}{\myArray[\x + 3*\y]}
    \node[fill=ltgray,circle, draw, minimum size=0.35cm , inner sep = 0pt] at (.6*\x,.6*\y) { $\j$}; 
     }
}
\end{scope}
\begin{scope}
\tiny
\draw[help lines] (0,0) grid[step=0.6cm, shift={(2.5,0)}] (1.2,1.2); 
\def\myArray{{2, 2, 1, 2, 2, 2, 1, 2, 2}}
\tiny
\foreach \x in {0,1,2}{                           
    \foreach \y in {0,1,2}{                       
    \pgfmathtruncatemacro{\j}{\x + 3*\y};
    \pgfmathtruncatemacro{\k}{\myArray[\j]};
    \node[fill=ltgray, draw, circle,minimum size=0.35cm , inner sep = 0pt](\j) at (2.5 + .6*\x,.6*\y) { $\k$}; 
    }
}
\draw[->] (6) edge (7) (7) edge (8) (8) edge (5) (5) edge (4) (4) edge (3) (3) edge (0) (0) edge (1) (1) edge (2);
\end{scope}
\begin{scope}
\draw[help lines] (0,0) grid[loosely dotted, step=0.6cm, shift={(5,0)}] (1.2,1.2); 
\def\myArray{{ 0, 1, 1, 1, 1, 1, 1, 1, 0}}
\tiny
\foreach \x in {0,1,2}{                         
    \foreach \y in {0,1,2}{                      
    \pgfmathtruncatemacro{\j}{\myArray[\x + 3*\y]}
    \node[fill=ltgray, draw, circle,minimum size=0.35cm , inner sep = 0pt] at (5+.6*\x,.6*\y) { $\j$}; 
    }
}
\end{scope}
\end{tikzpicture}
\caption{(left) Degrees of fermionic sites in the $3\times3$ 2D Hubbard lattice.  (middle) Degrees along the linear path $G_1$ through each sublattice (right). The nonlocal degree of each node determines the number of auxiliary modes to be used.}
\label{fig:edges}
\end{figure}

\section{Comparison}\label{sec:Hubb}
In the following we assess the various methods in context of the $2$D Hubbard Hamiltonian. We analyze locality of the nearest-neighbor fermionic hopping and density operators. The first subsection goes through locality analysis for $2$D Hubbard model. We then present results in higher dimensions. 

\subsection{2D Hubbard Model}
We consider a Hubbard model defined on a $w\times h, \; w\,< \,h$ rectangular lattice. For JW, we order the fermions as in Fig.~\ref{Fig:snake}. The longest string of $Z$ operators introduced by the mapping has then length $(w  + 1)$. 

For LSFS, we have already shown that the density-density terms in the Hubbard Hamiltonian are all $8$-local for lattices with $w,\, h \geq 3$. It is therefore only sensible to use LSFS for locality reduction if $w \geq 8$. Likewise AF only improves locality compared to JW if $w \geq 4$, since the most nonlocal term of the qubit Hamiltonian is the 4-local vertical hopping term.  

The AF transform is hence superior to LSFS for the Hubbard model defined on a rectangular lattice. 
Both methods use more qubits than fermionic sites of the original model. For a $w \times h$ rectangular lattice, LSFS requires $4(w-1)(h-1)$ qubits, while the auxiliary fermion method needs $4wh - 4$, compared to a minimum of $2wh$.  

We now compare these methods to the BK transform and its optimized SBK variant introduced in subsection~\ref{Subsec:Opti}. 
The density operator can be written in Majorana basis and converted by BK transform as:
\begin{align}
n_j &= \frac{\id + i c_j d_j } {2} \rightarrow \frac{1}{2} \left( \id - Z_{F(j)\cup \lbrace j \rbrace} \right) \,.
\end{align}
Its locality therefore only depends on $|F(j)|$, i.e. the number of children of a node $j$ in the Fenwick tree. The root has the largest number of children, which implies that the worst-case locality is $\left(\lfloor\log_2 N \rfloor + 1 \right)$. The density-density term is hence  $\left(2\lfloor\log_2 N \rfloor + 2 \right)$-local.

\begin{figure}[t]
\begin{tikzpicture}[every node/.style={minimum size =0.5cm}]

\node [rectangle, draw, thick, inner sep = 1pt] (15) at (0,0) {15};

\node [circle, draw, inner sep = 1pt, fill = black] (14) at (-2,-0.5) {\color{white} 14};
\node [circle, draw, inner sep = 1pt, fill = black] (13) at (-1,-.75) {\color{white} 13};
\node [circle, draw, inner sep = 1pt, fill = black] (11) at (0,-1) {\color{white} 11};
\node [rectangle, draw, inner sep = 1pt, fill = black] (7) at (2,-0.5) {\color{white} 7};

\node [circle, draw, inner sep = 1pt, fill = ltgray] (12) at (-1.5,-1.5) {12};
\node [circle, draw, inner sep = 1pt, fill = ltgray] (10) at (-.5, -2){10};
\node [circle, draw, inner sep = 1pt, fill = ltgray] (9) at (.5, -2){9};
\node [circle, draw, inner sep = 1pt, fill = ltgray] (6) at (1, -1){6};
\node [circle, draw, inner sep = 1pt, fill = ltgray] (5) at (2, -1.5){5};
\node [rectangle, draw, inner sep = 1pt, fill = ltgray] (3) at (3, -1){3};

\node [circle, draw, inner sep = 1pt, fill = ltgray] (8) at (0.5, -3){8};
\node [circle, draw, inner sep = 1pt, fill = ltgray] (4) at (2, -2.5){4};
\node [circle, draw, inner sep = 1pt, fill = ltgray] (2) at (3, -2){2};
\node [rectangle, draw, inner sep = 1pt, fill = ltgray] (1) at (4, -1.5){1};

\node [circle, thick, draw, inner sep = 1pt] (0) at (4, -2.5){0};

\draw (15) -- (14);
\draw (15) -- (13) -- (12);
\draw (15) -- (11) -- (10); 
\draw (11) -- (9) -- (8); 
\draw (15) -- (7); 
\draw (7) -- (6);
\draw (7) -- (5) -- (4);
\draw (7) -- (3) -- (2);
\draw (3) -- (1) -- (0);
\end{tikzpicture}
\caption{The worst case for hopping operator locality, if all-to-all couplings are allowed is the hopping from the first to the last node. In the diagram above, this corresponds to $0 \leftrightarrow 15$ - the nodes with raising/lowering operators are colored white. The set $F(15)$  of all children of node 15 corresponds to the black nodes.  The update set $U(0)$ of $0$ is the set of square nodes. This implies that $XZ = iY$ will be  applied at node 7.}
\label{Fig:WorstCase}
\end{figure}
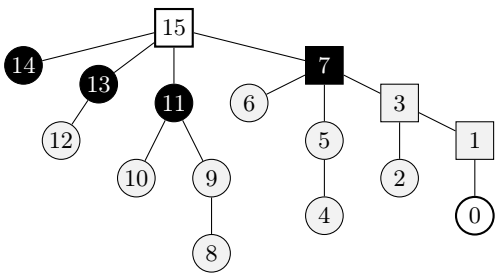

Using the expression for Majorana operators in the Bravyi-Kitaev mapping (Eq.~\ref{Eq:Majorana}), we have for the hopping operator that: 
\begin{align}
a_k^\dag a_j + a_j^\dag a_k &= \frac{i}{2} \left( c_k d_j +  c_j d_k \right)\,.
\end{align}
If all-to-all couplings were allowed, the worst-case locality would be $\lfloor \log_2 N \rfloor + \lceil \log_2 N \rceil$ - the sum of tree depth $d$ and the number of root children $n$. This corresponds to hopping from the first to the last site of the lattice (see Fig.~\ref{Fig:WorstCase}).

Focusing our attention to the nearest-neighbor hopping in the SBK transformation, we have to analyze the vertical and horizontal hoppings independently.  For hopping along horizontal edges, a loose bound on worst case locality is given by $2 \lceil \log_2 w \rceil -1 = 2d -1$ and corresponds to hopping to the root child with smallest index $j$ from the $(j+1)$-th node. 

\begin{figure}[th]
\includegraphics[scale=0.8]{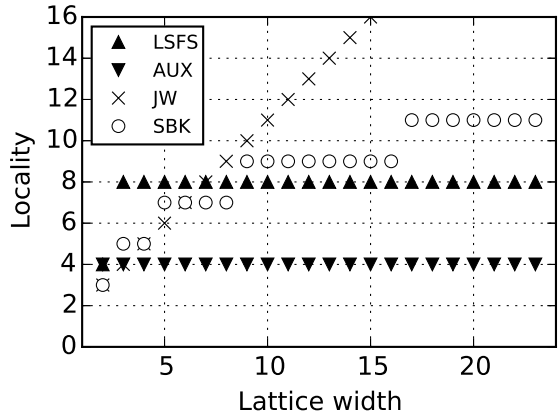}
\caption{Worst case locality of qubit operators with lattice size in the $2$D Hubbard model.} 
\label{Fig:summary1}
\end{figure}

The worst case locality for hopping along the verticals is given by $2 \lceil \log_2 w \rceil + 2 = 2d +2$ (where $d$ is the tree depth) and corresponds to hopping between two deepest leaf nodes. This can be reduced to $2 \lceil \log_2 w \rceil + 1  = 2 d+ 1$ by splitting the segmented trees in half, which however increases the worst case locality for horizontal hoppings to $2 \lceil \log_2 w \rceil$. The scheme hence provides locality advantage compared to JW for lattices of width $w > 2$. The worst case localities for $2$D are sumarized in Tab.~\ref{Tab:summary} and in Fig.~\ref{Fig:summary1}.

\squeezetable
\begin{table*}[tbh]
\begin{ruledtabular}
\begin{tabular}{ccccc}
Method & Density-density			& Horizontal				& Vertical		 & Qubits\\
\hline
JW     &         2          &      			2	 &		 w+1   	 & $2wh$ \\
BK   &      $2\lfloor \log_2 (wh)  \rfloor+ 2$             &         $\lfloor \log_2 (wh)  \rfloor  +\lceil \log_2 (wh)  \rceil $            &  $\lfloor \log_2 (wh)  \rfloor  +\lceil \log_2 (wh)  \rceil $             & $2wh$\\
SBK    & 	 $2\lfloor \log_2 w \rfloor + 2	$	& $\lfloor \log_2 w  \rfloor  +\lceil \log_2 w  \rceil $ 	      & 	$2 \lfloor \log_2 w \rfloor + 1$  		 & $2wh$ \\
AF   &  2           &   			2	 &  		4     & $4(wh-1)$  \\
LSFS    &  8 	   & 7  &   7         & $4wh - 2h - 2w $ 
\end{tabular}
\end{ruledtabular}
\caption{Operator locality and qubit resource overheads for transformation of a $2$D rectangular lattice Hubbard model. } 
\label{Tab:summary}
\end{table*}

\begin{table}[tbh]
\begin{ruledtabular}
\begin{tabular}{ccc}
Method & Worst-case locality & Qubits  \\
\hline
JW & $w^{D-1} + 1$ & $2w^D$  \\
BK & $2 \lfloor \log_2 w^D \rfloor$ & $2w^D$\\
SBK & $2 \lfloor \log_2 w^{D-1} \rfloor +1 $ & $2w^D$  \\
AF & $2D$ & $2D \, w^D$ \\
LSFS & $4D$ & $2D(w-1)w^{D-1} $
\end{tabular}
\end{ruledtabular}
\caption{Worst-case locality of the hopping term as a function of the hypercubic lattice dimension.}
\label{Tab:dimension}
\end{table}

\subsection{Hypercubic Lattices}
We now extend the previous analysis to higher dimensions and consider mapping of the hopping term in a general hypercubic lattice of $D > 0$ dimensions with side $w$. With the simplest ordering, the hopping term in Jordan-Wigner transformation becomes $w^{d-1} + 1$ local, where $w$ is the number of sites along a side of the hypercubic lattice.

In the case of AF, the number of auxiliaries  per site scales as $D-1$, since in $D$-dimensional hypercubic lattice, each vertex has $2D$ nearest neighbors. This translates to $d_{nl}=2D-2$ non-local degree and $D-1$ auxiliary fermion sites per each site of the original lattice. 
The worst case hopping term locality hence goes as $2D-2$, whereas the number of qubits scales as $D \times w^D$.  

Since each vertex has $2D$ neighbors, locality of the LSFS hopping term goes as $4D - 1$ in a bulk of the hypercube (one neighbor is shared), while locality of the density-density term becomes $4D$. 
The number of edges in a hypercubic lattice of side $w$ in dimension $D$ is given by $E(D, w) = D (w-1) w^{D-1} $, which is also the number of qubits required for the mapping. 

\begin{proof}
One has that $E(2, w) = 2(w-1)w$, as there are $2$ edges per vertex with the exception of the boundary, where we over-count by $2w$. In $3$ dimensions, we construct a $w\times w \times w$ cube by connecting vertices of $w\times w$ squares by $(w-1)w^2$ edges. It follows that $E(3, w) = wE(2,w) + w^2(w-1)  = 4 w^{2}(w-1)$. To count the number of edges in a $w^{\times D}$ hypercube, one can take $w$ $(D-1)$-dimensional hypercubes and connect them by $(w-1)w^{D-1}$ edges. This implies that $E(D, w) = wE(D-1, w) + w^{D-1} (w-1)$. It follows that $E(D, w) = D(w-1)w^{D-1}$.
\end{proof}

The BK method requires $N = w^D$ qubits in general, while its worst-case locality scales as $\lceil \log_2(w^D) \rceil  + \lfloor \log_2(w^D) \rfloor = d + n$. For nearest neighbour hopping, this can be further optimized to $2\lceil \log_2(w^{D-1})  \rceil+ 1 = 2d + 1$ by using the segmentation trick of the SBK method at the highest level of the lattice. Hopping term locality as a function of hypercubic lattice dimension is summarized in Tab.~\ref{Tab:dimension}. 

\section{Conclusion}

Which scheme is superior depends on the type of quantum simulation. For digital quantum simulation on a universal quantum computer with logical qubits, $N$-qubit operators can be simulated with just linear overhead in $N$. On such devices the main limitation will be the number of available logical qubits and then our  proposed modification of the Bravyi-Kitaev transform leads to the best locality improvement of $O(\log_2 w)$-local spin operators, where $w$ is the lattice width.

For analog simulations, on the other hand, operator locality will be the decisive factor. On such quantum simulators natively only few-qubit couplings are available, typically only two-qubit terms. In that setting multi-qubit terms have to be generated using perturbative gadgets \cite{KKR,Oliveira}. These, however, require large energy penalties to be sufficiently deep inside the perturbative regime where the effective higher-order interactions appear. One hence wants to optimize the locality of the terms and aim for transformations with the most local terms. The presented analysis shows that it is possible to map the $2$D Hubbard to a $4$-local qubit Hamiltonian by the Auxiliary Fermion method, at the expense of using number of ancillary qubits. This is the optimal fermion representation for 2D lattices of width $w\ge4$. 

\section{Acknowledgements} 
We would like to thank to Peter~D. Johnson, Peter Winkler, Amit Chakrabarti, Thomas~H. Cormen and Alexey Soluyanov for useful discussions. V.~H. would like to thank Dartmouth College for support and hospitality while finishing the work and the Clarendon and Keble de Breyne scholarships for support. This project was supported by
the Swiss National Science Foundation through the National
Competence Center in Research NCCR QSIT and by the European Research Council through ERC Advanced Grant SIMCOFE.

\bibliographystyle{unsrt}
\bibliography{operator_locality_in_qsim_of_fermionic_models}

\end{document}